\documentclass{article}

\usepackage{arxiv}
\usepackage[utf8]{inputenc} % allow utf-8 input
\usepackage[T1]{fontenc}    % use 8-bit T1 fonts
\usepackage{hyperref}       % hyperlinks
\usepackage{url}            % simple URL typesetting
\usepackage{booktabs}       % professional-quality tables
\usepackage{amsfonts}       % blackboard math symbols
\usepackage{nicefrac}       % compact symbols for 1/2, etc.
\usepackage{microtype}      % microtypography
\usepackage{lipsum}		% Can be removed after putting your text content
\usepackage{graphicx}
\usepackage{natbib}
\usepackage{doi}
\usepackage{float}

\title{Fully automated quantification of in vivo viscoelasticity of prostate zones using magnetic resonance elastography with Dense U-net segmentation}

\author{ {\hspace{1mm}Nader Aldoj}\thanks{Department of Radiology, Charité – Universitätsmedizin Berlin, Berlin, Germany} \\
	\texttt{nader.aldoj@charite.de} \\
	\And
	{\hspace{1mm}Federico Biavati} \footnotemark[1] \\
	\texttt{Federico.Biavati@charite.de} \\
	\And
	{\hspace{1mm}Marc Dewey} \footnotemark[1] \thanks{Berlin Institute of Health at Charité – Universitätsmedizin Berlin, Berlin, Germany} \thanks{DKTK (German Cancer Consortium), partner site Berlin, Germany}\\
	
	\texttt{marc.dewey@charite.de} \\
	\And
	{\hspace{1mm}Anja Hennemuth} \thanks{Institute of Computer-assisted Cardiovascular Medicine, Charité – Universitätsmedizin Berlin, Berlin, Germany}\\
	
	\texttt{Anja.Hennemuth@charite.de} \\
	\And
	{\hspace{1mm}Patrick Asbach} \footnotemark[1]\\
	
	\texttt{Patrick.Asbach@charite.de} \\
	\And
	{\hspace{1mm}Ingolf Sack} \footnotemark[1]\\
	
	\texttt{ingolf.Sack@charite.de} \\
}

\renewcommand{\shorttitle}{Learned prostate segmentation in MRE}

\hypersetup{
pdftitle={Fully automated quantification of in vivo viscoelasticity of prostate zones using magnetic resonance elastography with Dense U-net segmentation},
pdfsubject={q-bio.NC, q-bio.QM},
pdfauthor={Nader Aldoj, Ingolf Sack},
pdfkeywords={Magnetic resonance elastography, Multi-parmetric magnetic resonance imaging, prostate segmentation, prostate zones, neural networks, Dense U-net},
}

\begin{document}
\maketitle

\begin{abstract}
Magnetic resonance elastography (MRE) for measuring viscoelasticity heavily depends on proper tissue segmentation, especially in heterogeneous organs such as the prostate. Using trained network-based image segmentation, we investigated if MRE data suffice to extract anatomical and viscoelastic information for automatic tabulation of zonal mechanical properties of the prostate. 
Overall, 40 patients with benign prostatic hyperplasia (BPH) or prostate cancer (PCa) were examined with three magnetic resonance imaging (MRI) sequences: T2-weighted MRI (T2w), diffusion-weighted imaging (DWI), and MRE-based tomoelastography, yielding six independent sets of imaging data per patient (T2w, DWI, apparent diffusion coefficient (ADC), MRE magnitude, shear wave speed, and loss angle maps). Combinations of these data were used to train Dense U-nets with manually segmented masks of the entire prostate gland (PG), central zone (CZ), and peripheral zone (PZ) in 30 patients and to validate them in 10 patients. Dice score (DS), sensitivity, specificity, and Hausdorff distance were determined.
We found that segmentation based on MRE magnitude maps alone (DS, PG: 0.93±0.04, CZ: 0.95±0.03, PZ: 0.77±0.05) was more accurate than magnitude maps combined with T2w and DWI\_b (DS, PG: 0.91±0.04, CZ: 0.91±0.06, PZ: 0.63±0.16) or T2w alone (DS, PG: 0.92±0.03, CZ: 0.91±0.04, PZ: 0.65±0.08). Automatically tabulated MRE values were not different from ground-truth values (P>0.05). In conclusion, MRE combined with Dense U-net segmentation allows tabulation of quantitative imaging markers without manual analysis and independent of other MRI sequences and can thus contribute to PCa detection and classification.

\end{abstract}

% keywords can be removed

\section{Introduction}
\begin{figure}[ht]
\centering
\includegraphics[width=\linewidth]{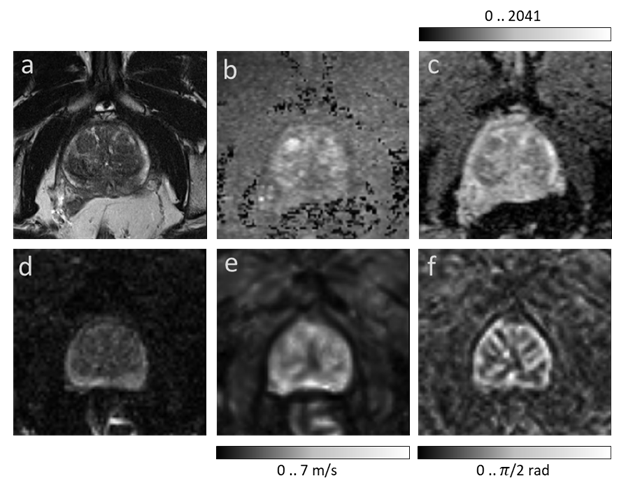}
\caption{Images illustrating the appearance of the prostate in mpMRI in a volunteer with benign prostate hyperplasia (BPH). (a) T2-weighted MRI, (b) DWI image with a b-value of 1400 s/mm2, (c) ADC map, (d) MRE magnitude signal, (e) SWS map as a surrogate marker of tissue stiffness, and (f) $\varphi$ map, which signifies friction or tissue fluidity.}
\label{fig:1.png}
\end{figure}

Prostate cancer (PCa) has the highest incidence of all types of cancer in men and is the second leading cause of cancer deaths in men \cite{RN127, RN135}. Radiological imaging modalities such as magnetic resonance imaging (MRI) play a central role in the diagnosis of PCa and therapy planning. In particular, multiparametric MRI as defined in the Prostate Imaging Reporting and Data System (PI-RADS) has contributed to the standardization of prostate MRI worldwide \cite{RN148}. However, inter- and intrareader agreement is still moderate \cite{RN149}. 
This is mainly due to the use of subjective imaging criteria in the assessment of lesion shape and signal intensity as well as manual segmentation procedures of prostate regions, resulting in high variability of reference values. It is noteworthy that even quantitative imaging parameters can vary widely with the size and site of the regions selected for their analysis \cite{RN150}. 
Therefore, accurate segmentation of the prostate gland and its zones is critical for the detection and management of PCa \cite{RN119, RN101, RN99, RN103}. Automated prostate segmentation using multiparametric MRI (mpMRI) is often performed based on morphologic images such as T2-weighted MRI and transferred to other images which depict microstructural information such as the apparent diffusion coefficient (ADC) measured by diffusion-weighted imaging (DWI). Complementary to DWI, magnetic resonance elastography (MRE) \cite{RN151} has been recently introduced for the clinical assessment of PCa \cite{RN152, RN142, RN157, RN171, RN169, RN170, RN168, RN172, RN173, RN174, RN177, RN178}. MRE provides maps of stiffness and viscosity, which are quantitatively linked with mechanical microstructures in biological soft tissues and their changes due to disease \cite{RN153, RN154}. Unlike DWI, where ADC is reconstructed from variations in signal magnitude, MRE values are reconstructed from phase images, while magnitude images have been largely unused except for recent approaches to water diffusion analysis \cite{RN155, RN156}. Our hypothesis was that MRE magnitude images provide anatomical information that can be used for automated segmentation of prostate zones. Full exploitation of anatomic and viscoelastic information contained in a single set of MRE data would greatly facilitate quantitative parameter extraction without co-registration artifacts. However, for several reasons, automated prostate segmentation is a challenging task \cite{RN112}. For example, the prostate is a highly heterogeneous organ with complex 3D geometry giving rise to well-known ambiguities of tissue boundaries in MRI. Furthermore, prostate morphology greatly varies among individuals, especially when benign prostatic hyperplasia (BPH) or advanced PCa is present. Figure \ref{fig:1.png} illustrates these challenges in a representative case of BPH.
Numerous studies have tackled the problem of automatic segmentation of MR images of the prostate using various approaches such as atlas segmentation \cite{RN106}, deformable and statistical modeling \cite{RN131} or machine learning \cite{RN134}. %\cite{zheng2014marginal}.
For a few years, deep convolutional neural networks (CNNs) have been extensively used for segmentation tasks in various radiological applications, thanks to their suitability for generalization \cite{RN133}. In the prostate, CNNs were used for prostate gland segmentation using either slice-wise \cite{RN126} or full 3D approaches \cite{RN113, RN116} However, only a few approaches were actually able to segment the prostate gland (PG)  and its zones such as the central zone (CZ) and the peripheral zone (PZ) \cite{RN98, RN105, RN132, RN109}.

In this study, we used CNNs for automated segmentation of the prostate zones based on T2-weighted (T2w) MRI, DWI, and MRE \cite{RN146}. DWI provides intensity images (based on a specific b-value, DWI\_b) and quantitative ADC maps. 
MRE provides magnitude images (mag), shear wave speed maps of stiffness (SWS, in m/s), and maps of loss angle ($\varphi$ in radians, indicating friction or a material’s fluidity) \cite{RN179}. We explored CNNs for prostate segmentation using 14 possible input combinations of MRE maps as input training data. First, we trained and tested individual networks (models), henceforth termed individual models (IMs), using each of the aforementioned sets with their corresponding manually segmented masks of CZ, PZ, and PG as ground truths. Second, the dataset was rearranged so that it contains all 14 possible input combinations at once to be used as input training data for a single model, henceforth termed unified model (UM). 
The purpose of this study was to test whether MRE-based tomoelastography data are sufficient to extract anatomic and viscoelastic information for automatic tabulation of zonal mechanical properties of the prostate and to compare MRE-based image segmentation with information extracted by CNNs from other MRI pulse sequences typically acquired in clinical practice. 

\section{Methods}
\label{sec:headings}

\subsection{Subjects}

The imaging data used in this study were used retrospectively and were acquired in a previously reported population of BPH and PCa patients who underwent PI-RADS-compatible mpMRI and MRE \cite{RN142}. Our local ethical review board approved this study, and all patients gave written in-formed consent. Forty patients were included, 26 with a PI-RADS score of 2 consistent with BPH. Fourteen men had a suspicious focal lesion (PI-RADS 4: n = 2, maximum tumor diameters of 10 and 11 mm; PI-RADS 5: n = 12, maximum tumor diameters ranging from 16 to 66 mm). In these 14 men, prostate biopsy revealed PCa Gleason scores of 3 + 3 (ISUP class 1, n = 1), 3 + 4 (ISUP class 2, n = 2), 4 + 3 (ISUP class 3, n = 2), 4 + 4 (ISUP class 4, n = 4), and $\geq$ 4 + 5 (ISUP class 5, n = 5).

From the 40 volume datasets of these patients, 30 were randomly selected for training. Each vol-ume dataset included 25 slices, yielding a total of 25 (slices) x 30 (volumes) x 1 (input combination) = 750 training images for each of the 14 input combinations of IMs, and 25 (slices) x 30 (volumes) x 14 (all input combinations) = 10,500 images for UM. In the latter case, all 14 input combinations were included in a single large training set. The remaining 10 datasets with a total of 25 (slices) x10 (volumes) = 250 slices were used for testing.

\subsection{MR Imaging}
The imaging data were acquired on a 3-Tesla MRI scanner (Magnetom Skyra; Siemens Healthi-neers, Erlangen, Germany) using both the 18-channel phased-array surface coil and a spine array coil. All patients underwent a clinically indicated mpMRI examination of the prostate in accord-ance with PI-RADS version 2 (2015) \cite{RN158} which included T2-weighted sequences (T2w) in the axial and coronal planes and DWI. Here we used b-values of 0, 50, 500, 1000, and 1400 s/mm2 (b=1400 s/mm2 was used as diffusion weighted image, henceforth referred to as DWI\_b). ADC maps were automatically generated by the MRI scanner via monoexponential fitting of all b-values. After the clinical MRI examination, patients underwent multifrequency MRE-based tomo-elastography using a single-shot spin-echo sequence with three excitation frequencies of 60, 70, 80 Hz \cite{RN142}. MRE magnitude images (mag) display signal intensities which are T2-weighted with a strong T2* effect. All imaging parameters are summarized in Table \ref{tab:table1}.

%----------------------------Table 1 ---------------------------
\begin{table}
	\caption{Imaging parameters of MRE and MRI}
	\centering
	\resizebox{\columnwidth}{!}{%
	\begin{tabular}{c|cccccccc}
		\toprule
			Map & \shortstack{Axial slices \\ (number) } & \shortstack{Slice thickness \\ ($mm$)} & \shortstack{Field of view \\ ($mm^2$)}  & \shortstack{Matrix size\\ (pixels)} & \shortstack{Voxel size\\ ($mm^3$)} 	& \shortstack{Repetition time (TR)\\ ($ms$)} & \shortstack{Echo time (TE) \\ ($ms$)}  & \shortstack{Parallel imaging factor \\ (GRAPPA-algorithm)}  \\
		\midrule
		MRE & 25 & 2 $mm$ & 256x256 $mm^2$ & 128x128 & 2x2x2 $mm^3$ & 3240 $ms$ & 69 $ms$ &  2     \\
		T2w &    25 & 3 $mm$ & 180x180 $mm^2$ & 384x384 & 0.47x0.47x3 $mm^3$	& 4000 $ms$ & 116 $ms$ & 2      \\
		DWI &   25 & 3 $mm$ & 230x230 $mm^2$ & 160x160 & 1.44x1.44x3 $mm^3$	& 4000 $ms$ & 56 $ms$ &  2  \\
		\bottomrule
	\end{tabular}
	\label{tab:table1}
	}
\end{table}

MRE data processing was based on multifrequency wavefield inversion for generating frequency-compounded SWS and $\varphi$ maps. SWS was reconstructed by single-derivative finite-difference operators (k-MDEV) \cite{RN159} while $\varphi$ was obtained by second-order, Laplacian-based, direct inversion (MDEV) \cite{RN160}. Although k-MDEV is noise-resistant and well-suited for SWS visualization, it is limited regarding the quantification of viscosity-related parameters such as $\varphi$ \cite{RN159}. Hence, MDEV was used for $\varphi$ recovery. Both k-MDEV and MDEV pipelines are publicly available at bioqic-apps.charite.de \cite{RN161}. Two experienced radiologists (PA with more than 10 years and FB with 2 years of experience in PIRADS-based PCa detection and classification) segmented and revised the prostate and its zones based on T2w-MRI, DWI, and MRE. The resulting masks of CZ (which includes the transition zone), PZ, and PG were used as ground truth (GT) for training and validation of CNNs. Overall, six independent imaging data were further used for segmentation analysis: T2w, DWI\_b, ADC, mag, SWS, and $\varphi$.

\subsection{Image preparation and augmentation}
\begin{figure}[ht]
\centering
\includegraphics[width=\linewidth]{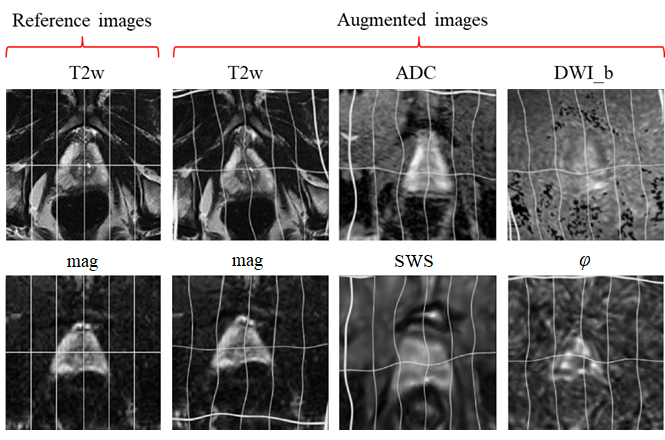}
\caption{Image augmentation using elastic deformation. First column shows the reference non-augmented T2w images in the first row and MRE magnitude maps (mag) in the second row. Augmented images are shown in the remaining columns: first row: T2w, ADC, and DWI\_b. Second row: mag, SWS and $\varphi$.}
\label{fig:2.png}
\end{figure}

Since T2w, DWI, and MRE images had different resolutions, all images were resampled to a common resolution of 0.5 mm isotropic edge length. Images of the same size and resolution were obtained by positioning a cropping window with a size of 256x256 pixels at the center of each 3D imaging volume. For image augmentation, 9 random elastic deformations were applied to the original images to increase the number of training sets. As described in \cite{RN110}, elastic deformation can be driven by two main parameters: $\sigma$, which represents the elasticity coefficient, and $\alpha$, which represents a scaling factor that controls the amplitude of the deformation. The two parameters were set to $\alpha$ = 21 and $\sigma$ = 512. Examples of augmented images are provided in Figure \ref{fig:2.png}.

\subsection{Dense U-net}
\begin{figure}[ht]
\centering
\includegraphics[width=\linewidth]{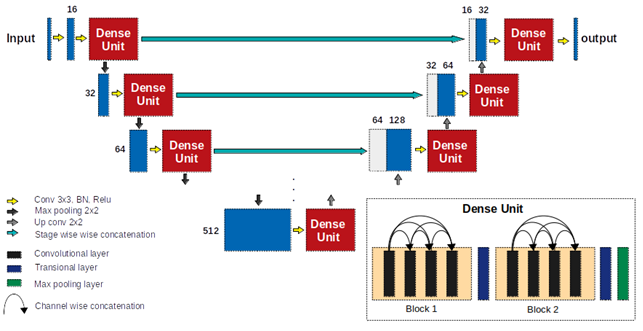}
\caption{Dense U-net architecture. Numbers in the figure indicate the number of feature maps at each stage.}
\label{fig:3.png}
\end{figure}

A previously developed Dense U-net \cite{RN146} based on state-of-the-art U-net convolutional network architecture \cite{RN132} was used. The network comprised two main parts - an encoder and a decoder - between which skip connections connected feature maps with similar resolutions. Normal stacks of convolutional layers at each stage were substituted with two densely connected blocks. Each dense block consisted of four convolutional layers with 3x3 kernel size followed by a transitional layer. The structure of the network is illustrated in Figure \ref{fig:3.png}.

\subsection{Network training}
Stochastic gradient descent with a learning rate of 10-3, a momentum of 0.9, and a decay of 10-6 was used in this study to train and test all proposed models. We used cross-entropy (CE) loss as the main loss function, and performed pixel-wise comparison of ground-truth images and the resulting masks. This loss was represented by

\begin{equation}
 loss_{CE}(m,I) = - (m log (I) + (1-m) log(1-I)) 
\end{equation}

with \textit{m} and \textit{I} denoting the prediction masks and the ground-truth images, respectively.

\begin{figure}[ht]
\centering
\includegraphics[width=\linewidth]{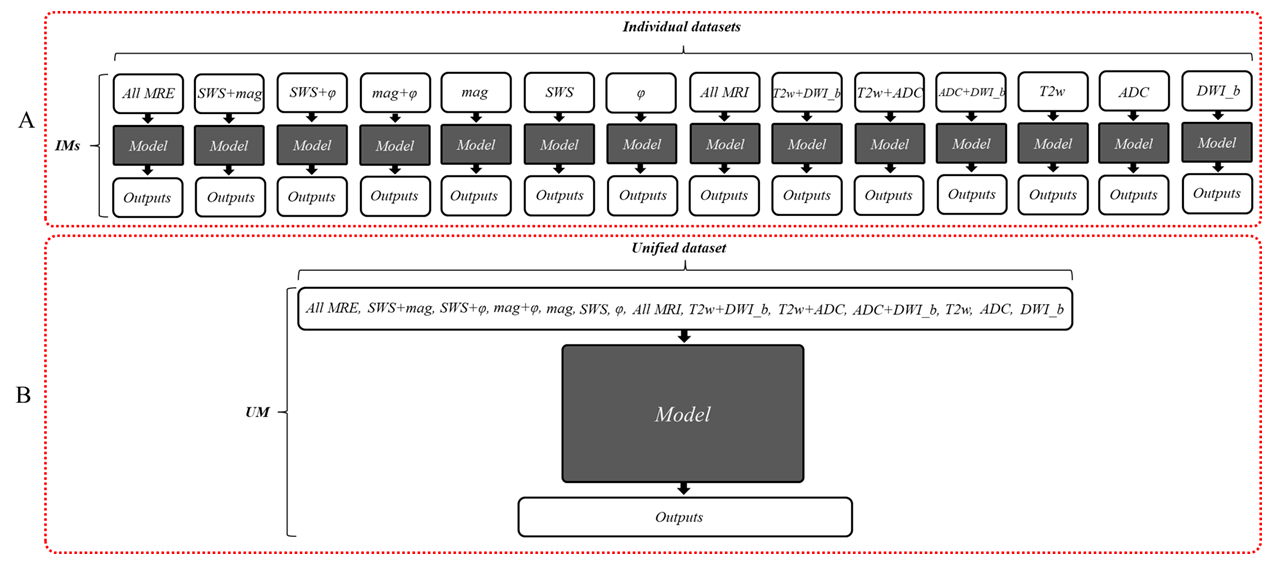}
\caption{Diagrams of the two approaches investigated in this study. (A) individual models (IMs), and (B) unified model (UM). The figure shows all 14 input combinations used individually as separate datasets in IMs and combined into a single large dataset for training of the UM. The gray box with the name ‘Model’ refers to the Dense U-net used in this study, as detailed in Figure \ref{fig:3.png}}
\label{fig:4.png}
\end{figure}

Two main approaches were tested: 
(i) Individual models (IMs), where we trained and tested a separate model for each combination of sequences/maps or individual sequence/map input.
(ii) Unified model (UM) without MRE-MRI co-registration (i.e., a single model for T2w, DWI\_b, ADC, mag, SWS, and $\varphi$ images, trained on masks that were manually and separately segmented from T2w and mag images), yielding a re-arranged dataset, where all image combinations were taken into account during training, validation and testing.
Fourteen input combinations were used for training and testing IMs and the UM, more specifically mag+SWS+$\varphi$, SWS+mag, SWS+$\varphi$, mag+$\varphi$, mag, $\varphi$, SWS, T2w+ADC+DWI\_b, T2w+ADC, T2w+ADC, ADC+DWI\_b, T2w, ADC and DWI\_b. While each of these input combinations was used to train an individual model in IMs, in UM, each input combination represented a subset and was combined with all other subsets to generate a single, large dataset of all 14 input combinations for training and testing the UM. Figure \ref{fig:4.png} shows how imaging data were used as inputs for IMs and the UM.

\subsection{Evaluation}
We evaluated all resulting segmentations against manually delineated ground-truth masks using standard evaluation statistics such as mean dice score (DS) ± standard deviation (SD), sensitivity (Sen), specificity (Spc), and Hausdorff distance (HD) as a contour consistency measure, and a t-test with p<0.05 indicating a statistically significant difference.

The dice score is a similarity measure that quantifies the overlap between predicted masks and ground-truth labels, allowing straightforward comparison of segmentation performance \cite{RN123}. DS was formulated for two binary sets, \textit{A} and \textit{B}, as follows:

\begin{equation}
  DS(A,B) = \frac{2|A \cap B|}{|A| + |B|} 
\end{equation}

The Hausdorff distance is the maximum distance between two edge points from two different sets (predicted mask and ground truth). It is expressed in millimeters (mm) and was defined as follows: 

\begin{equation}
  HD(A, B) = max \{max_{i\in A} min_{j\in B} d(i,j), max_{i\in B} min_{j\in A} d(i,j) \} 
\end{equation}

where \textit{d(i,j)} is the Euclidean distance between two points from different sets \textit{A} and \textit{B}.

\subsection{Implementation details}
All models were trained on individual or combinations of MRI/MRE images in a slice-wise fashion, where all images had a size of 256x256 and an in-plane resolution of 0.5x0.5 mm. We used the SimpleITK library for image preprocessing \cite{RN164, RN165}, and Keras with Tensorflow back-end \cite{RN166} as the main library for model implementation, training, and testing. All models were trained on a TitanXP GPU with 2 GB video memory (CUDA version of 10.1) and a batch size of 25 images. Training time of the Dense U-net was around 8.5 hours. The computation time during testing for a single 3D volume (of around 25 slices) was approximately 1.5 sec.

%\paragraph{Paragraph}

\section{Results}
\begin{figure}[ht]
\centering
\includegraphics[width=\linewidth]{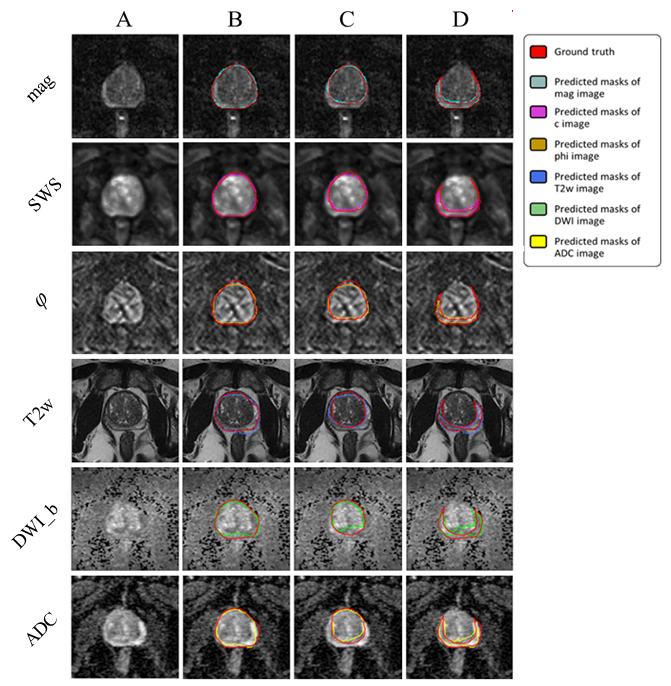}
\caption{Examples of segmentation results achieved with IMs: first column (A) shows the original image, second (B), third (C) and forth column (D) show masks of prostate, central and peripheral zones, respectively.}
\label{fig:5.png}
\end{figure}

\begin{figure}[ht]
\centering
\includegraphics[width=\linewidth]{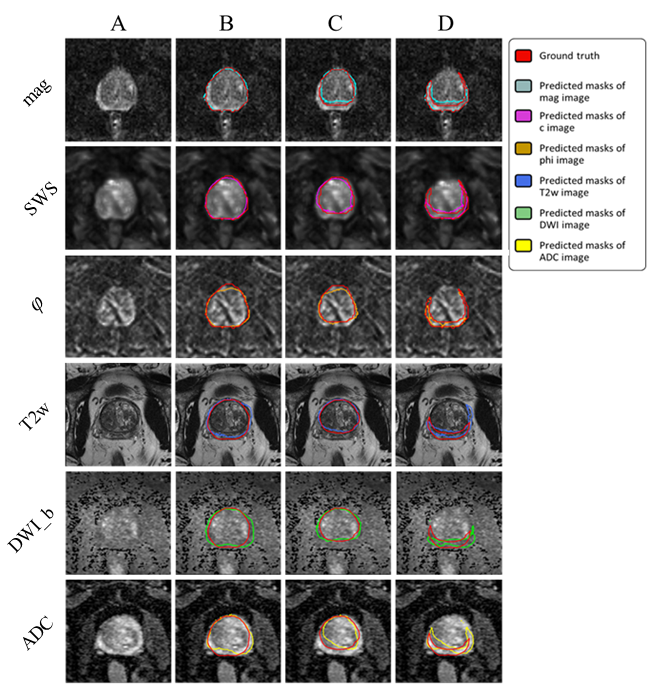}
\caption{Examples of segmentation results achieved with the UM: first column (A) shows the original image, second (B), third (C) and forth column (D) show masks of prostate, central and peripheral zones, respectively.}
\label{fig:6.png}
\end{figure}

Figure \ref{fig:5.png} and Figure \ref{fig:6.png} illustrate segmentation results for each of the two training approaches. The boundary of automatically segmented regions closely follows the boundary of manual segmentations.

Our experimental results for IMs trained on different combinations of maps showed that the proposed method was capable of efficiently segmenting PG, CZ, and PZ from the input MRI/MRE images. DS ranged from 0.87±0.04 (ADC) to 0.93±0.04 (mag), from 0.85±0.09 (DWI\_b) to 0.95±0.03 (mag), and from 0.53±0.10 (ADC+DWI\_b) to 0.77±0.05 (mag), for PG, CZ, and PZ, respectively. Also, HD was minimized (i.e., best contour consistency) based on mag with 0.86, 0.76, 0.98 mm (Figure \ref{fig:5.png} and Table \ref{tab:table2}). However, performance differences between MRE (mag, SWS, $\varphi$) and MRI maps (T2w, DWI\_b, ADC) were statistically not significant except for CZ, where MRE (mag, SWS, $\varphi$) had a significantly higher DS than MRI (T2w, DWI\_b, ADC, p<0.05).

%-------------------------------Table 2 --------------------------------

\begin{table}
	\caption{Statistical analysis of the segmentation results of IMs. Bold indicates the highest values, while underlining indicates the lowest values. ↑ (↓): the higher (the lower) the better.}
	\centering
	\resizebox{\columnwidth}{!}{%
	\begin{tabular}{l|cccccc|cccccc|cccccc}
		\toprule
		\multicolumn{6}{c}{\hspace{3cm}Prostate gland} &
		\multicolumn{6}{c}{\hspace{2cm}Central zone} &
		\multicolumn{6}{c}{\hspace{2cm}Peripheral zone}\\
		\cmidrule(r){2-7}
		\cmidrule(r){8-13}
		\cmidrule(r){14-19}
		Model &	Dice ↑ & Std ↓ & Median ↑ & Sen ↑ & Spc ↑ & HD ↓ & Dice ↑ & Std ↓ & Median ↑ & Sen ↑ & Spc ↑ & HD ↓ & Dice ↑ & Std ↓ & Median ↑ & Sen ↑ & Spc ↑ &HD ↓ \\
		\midrule
		All MRE & 0.92 & 0.04 & 0.91 &	0.92 &	0.99 &	1.04 &	0.94 &	0.03 &	0.93 &	0.93 &	0.99	 & 1.08 & 0.73 & 0.05 &	0.74 &	0.72 &	1.00 &	1.35 \\
		SWS+mag & 0.91 & 0.04 &	0.90 &	0.90 &	0.99 &	1.22 &	0.93 &	0.04 &	0.92 &	0.91 &	0.99	 & 1.41 & 0.69 & 0.04 &	0.69 &	0.65 &	1.00 &	1.72      \\
		SWS+$\varphi$ & 0.91 & 0.05 &	0.90 &	0.89 &	0.99 &	1.12 &	0.92 &	0.05 &	0.93 &	0.92 &	0.99	 & 1.10 & 0.64 & 0.10 &	0.62 &	0.68 &	1.00 &	1.43  \\
		
		Mag+$\varphi$ & 0.91 & 0.04 & 0.91 & 0.94 & 0.99 & 1.95 & 0.93 & 0.04 & 0.93 & 0.95 & 0.99 & 1.92 & 0.69 & 0.06 & 0.68 & 0.68 & 1.00 & 1.97 \\
		Mag & \textbf{0.93} & 0.04 & 0.93 & 0.93 & 0.99 & \textbf{0.86} & \textbf{0.95} & 0.03 & 0.95 & 0.96 & 0.99 & \textbf{0.76} & \textbf{0.77} & 0.05 & 0.79 & 0.79 & 1.00 & \textbf{0.98}     \\
		$\varphi$ & 0.90 & 0.05 & 0.88 & 0.87 & 0.99 & 1.18 & 0.91 & 0.05 & 0.92 & 0.89 & 0.99 & 1.19 & 0.52 & 0.11 & 0.54 & 0.57 & 1.00 & 1.66  \\
		
		SWS & 0.82 & 0.08 & 0.80 & 0.78 & 0.99 & \underline{2.41} & 0.88 & 0.06 & 0.87 & 0.84 & 0.99 & \underline{2.56} & 0.55 & 0.07 & 0.53 & 0.59 & 1.00 & 2.64 \\
		All MRI & 0.90 & 0.02 & 0.88 & 0.83 & 1.00 & 1.06 & 0.88 & 0.05 & 0.86 & 0.82 & 1.00 & 1.16 & 0.66 & 0.11 & 0.67 & 0.61 & 1.00 & 1.81      \\
		T2w+ADC & 0.90 & 0.04 & 0.89 & 0.88 & 0.99 & 1.30 & 0.87 & 0.06 & 0.88 & 0.84 & 1.00 & 1.35 & 0.62 & 0.13 & 0.66 & 0.65 & 0.99 & 2.82  \\
		
		T2w+DWI\_b & 0.91 & 0.04 & 0.90 & 0.90 & 1.00 & 0.98 & 0.91 & 0.06 & 0.90 & 0.86 & 1.00 & 1.00 & 0.63 & 0.16 & 0.66 & 0.68 & 0.99 & 2.11 \\
		ADC+DWI\_b & 0.88 & 0.05 & 0.87 & 0.84 & 0.99 & 1.44 & 0.87 & 0.06 & 0.86 & 0.83 & 0.99 & 1.46 & \underline{0.53} & 0.10 & 0.55 & 0.52 & 0.99 & 2.84\\
		T2w & 0.92 & 0.03 & 0.92 & 0.93 & 0.99 & 1.05 & 0.91 & 0.04 & 0.91 & 0.89 & 1.00 & 0.87 & 0.65 & 0.08 & 0.65 & 0.72 & 0.99 & \underline{2.90}\\
		
		ADC & \underline{0.87} & 0.04 & 0.86 & 0.86 & 0.99 & 1.41 & 0.86 & 0.06 & 0.84 & 0.84 & 0.99 & 1.41 & 0.55 & 0.08 & 0.57 & 0.55 & 0.99 & 2.38\\
		DWI\_b & 0.88 & 0.06 & 0.86 & 0.87 & 0.99 & 1.38 & \underline{0.85} & 0.09 & 0.83 & 0.82 & 1.00 & 1.48 & 0.54 & 0.17 & 0.51 & 0.61 & 0.99 & 2.07\\
		\midrule
		All MRE ave. & 0.91 & 0.04 & 0.90 & 0.91 & 0.99 & 1.23 & 0.93 & 0.04 & 0.93 & 0.93 & 0.99 & 1.24 & 0.67 & 0.07 & 0.67 & 0.68 & 1.01 & 1.52\\
		All MRI ave. & 0.89 & 0.04 & 0.88 & 0.87 & 0.99 & 1.23 & 0.88 & 0.06 & 0.87 & 0.84 & 1.00 & 1.25 & 0.60 & 0.12 & 0.61 & 0.62 & 0.99 & 2.42\\
		\bottomrule
	\end{tabular}
	\label{tab:table2}
}
\end{table}

Unlike IMs, the UM can process any map or combination of maps, which facilitates clinical applications. Compared with IMs, the UM had higher DS in PG and CZ and lower DS in PZ. Assessed for different prostate zones, DS of the UM ranged from 0.77±0.11 (DWI\_b) to 0.92±0.04 (mag, SWS, $\varphi$), from 0.65±0.08 (DWI\_b) to 0.86±0.06 (mag, SWS, $\varphi$), and from 0.28±0.10 (DWI\_b) to 0.57±0.05 (mag), for PG, CZ, and PZ, respectively. The smallest HD of 1.15, 1.45, and 1.81 for PG, CZ, and PZ, respectively, was found for mag. Images are presented in Figure \ref{fig:6.png}, and the results are summarized in Table \ref{tab:table3}.

%---------------------------------Table 3 ----------------------------------

\begin{table}
	\caption{Statistical analysis of the segmentation results of the UM. Bold indicates the highest values, while underlining indicates the lowest values. ↑ (↓): the higher (the lower) the better. }
	\centering
	\resizebox{\columnwidth}{!}{%
	\begin{tabular}{l|cccccc|cccccc|cccccc}
		\toprule
		\multicolumn{6}{c}{\hspace{3cm}Prostate gland} &
		\multicolumn{6}{c}{\hspace{2cm}Central zone} &
		\multicolumn{6}{c}{\hspace{2cm}Peripheral zone}\\
		\cmidrule(r){2-7}
		\cmidrule(r){8-13}
		\cmidrule(r){14-19}
		Model &	Dice ↑ & Std ↓ & Median ↑ & Sen ↑ & Spc ↑ & HD ↓ & Dice ↑ & Std ↓ & Median ↑ & Sen ↑ & Spc ↑ & HD ↓ & Dice ↑ & Std ↓ & Median ↑ & Sen ↑ & Spc ↑ &HD ↓ \\
		\midrule
		All MRE & \textbf{0.92} & 0.04 & 0.92 & 0.87 & 1.00 & 1.17 & 0.86 & 0.06 & 0.82 & 0.80 & 0.99 & 1.61 & \textbf{0.56} & 0.05 & 0.56 & 0.54 & 0.99 & 2.00 \\
		
		SWS+mag & 0.91 & 0.04 & 0.91 & 0.87 & 1.00 & 1.19 & 0.85 & 0.06 & 0.81 & 0.78 & 0.99 & 1.64 & 0.55 & 0.05 & 0.56 & 0.54 & 0.99 & 2.01      \\
		SWS+$\varphi$ & 0.90 & 0.04 & 0.89 & 0.87 & 0.99 & 1.41 & 0.85 & 0.05 & 0.83 & 0.79 & 0.99 & 1.93 & 0.49 & 0.06 & 0.52 & 0.53 & 0.99 & 2.16  \\
		Mag+$\varphi$ & 0.91 & 0.04 & 0.91 & 0.88 & 0.99 & 1.23 & 0.87 & 0.05 & 0.85 & 0.84 & 0.99 & 1.66 & 0.57 & 0.03 & 0.57 & 0.55 & 0.99 & 1.86 \\
		
		Mag & 0.91 & 0.05 & 0.91 & 0.87 & 0.99 & \textbf{1.15} & \textbf{0.87} & 0.05 & 0.84 & 0.84 & 0.99 & \textbf{1.45} & 0.57 & 0.05 & 0.57 & 0.52 & 0.99 & \textbf{1.81}\\
		$\varphi$ & 0.88 & 0.04 & 0.88 & 0.85 & 0.99 & 1.54 & 0.83 & 0.04 & 0.82 & 0.76 & 0.99 & 2.03 & 0.44 & 0.07 & 0.47 & 0.49 & 0.98 & 2.43\\
		SWS  & 0.89 & 0.05 & 0.89 & 0.87 & 0.99 & 1.63 & 0.82 & 0.08 & 0.80 & 0.76 & 0.99 & 2.18 & 0.46 & 0.06 & 0.47 & 0.48 & 0.99 & 2.19\\
		
		All MRI & 0.80 & 0.09 & 0.78 & 0.91 & 0.98 & 2.37 & 0.73 & 0.06 & 0.71 & 0.56 & 1.00 & 2.80 & 0.51 & 0.09 & 0.53 & 0.56 & 0.99 & 3.42\\
		
		T2w+ADC & 0.80 & 0.09 & 0.78 & 0.91 & 0.98 & 2.34 & 0.72 & 0.05 & 0.70 & 0.55 & 1.00 & 2.86 & 0.49 & 0.09 & 0.52 & 0.55 & 0.99 & 3.56\\
		T2w+DWI\_b & 0.81 & 0.09 & 0.80 & 0.90 & 0.98 & 2.04 & 0.74 & 0.06 & 0.72 & 0.58 & 1.00 & 2.63 & 0.40 & 0.08 & 0.41 & 0.41 & 0.99 & 3.82\\
		ADC+DWI\_b & 0.80 & 0.09 & 0.75 & 0.88 & 0.98 & 2.25 & 0.670 & .07 & 0.64 & 0.48 & 1.00 & 3.45 & 0.45 & 0.11 & 0.44 & 0.53 & 0.99 & 4.10\\
		
		T2w & 0.78 & 0.09 & 0.78 & 0.87 & 0.98 & \underline{2.43} & 0.75 & 0.07 & 0.72 & 0.60 & 1.00 & 2.35 & 0.42 & 0.12 & 0.43 & 0.38 & 0.99 & 4.12\\
		ADC & 0.80 & 0.08 & 0.76 & 0.88 & 0.98 & 2.21 & \underline{0.65} & 0.07 & 0.62 & 0.46 & 1.00 & \underline{3.54} & 0.43 & 0.11 & 0.43 & 0.52 & 0.98 & 4.19\\
		DWI\_b & \underline{0.77} & 0.11 & 0.76 & 0.88 & 0.98 & 2.42 & 0.65 & 0.08 & 0.64 & 0.49 & 1.00 & 3.38 & \underline{0.28} & 0.10 & 0.30 & 0.33 & 0.98 & \underline{5.29}\\
		
		\midrule
		All MRE ave. & 0.89 & 0.05 & 0.89 & 0.87 & 0.99 & 1.46 & 0.83 & 0.06 & 0.81 & 0.77 & 0.99 & 1.91 & 0.52 & 0.06 & 0.53 & 0.53 & 0.99 & 2.23\\
		All MRI ave. & 0.79 & 0.09 & 0.77 & 0.89 & 0.98 & 2.29 & 0.70 & 0.07 & 0.68 & 0.53 & 1.00 & 3.00 & 0.43 & 0.10 & 0.44 & 0.47 & 0.99 & 4.07\\
		\bottomrule
	\end{tabular}
	\label{tab:table3}
}
\end{table}

\begin{figure}[ht]
\centering
\includegraphics[width=\linewidth]{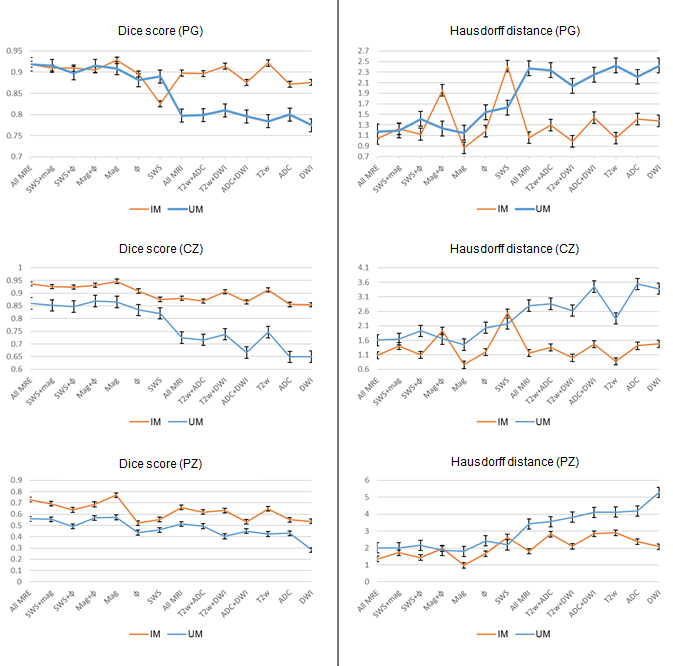}
\caption{Comparison of the performance of the two approaches: IMs with orange and UM with blue color. The left and right figures show the values of dice score and Hausdorff distance, respectively.}
\label{fig:7.png}
\end{figure}

\begin{figure}[ht]
\centering
\includegraphics[width=\linewidth]{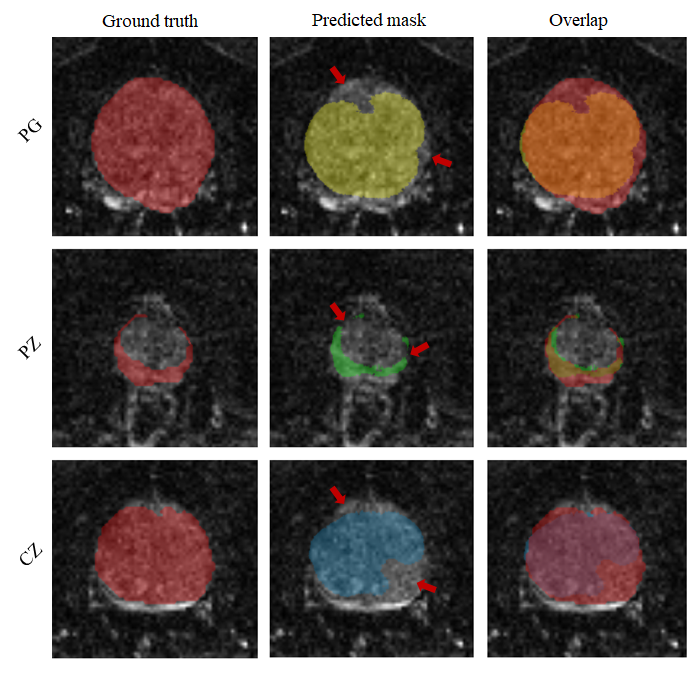}
\caption{Illustration of inaccurate segmentations: left column is the ground truth; middle is predicted masks and right is the overlap between ground truth and predicted masks. Rows from top to bottom, images of PG, PZ and CZ respectively.}
\label{fig:8.png}
\end{figure}

Unlike IMs, the UM showed a significantly higher DS when using MRE data in comparison with MRI data (p < 0.001, 0.05, and 0.05 for PG, CZ, and PZ, respectively).
Overall, IMs had significantly more accurate results compared with the UM in terms of both DS and HD with p-values < 0.01 for PG, CZ, and PZ. This is shown in Figure \ref{fig:7.png}, illustrating that DS was higher and HD lower for IMs than for the UM. 
Figure \ref{fig:8.png} shows a case where model segmentations were inaccurate compared with the ground-truth masks. Quantitative analysis, for both IMs and the UM, of pixel values in PG, CZ, and PZ showed no significant difference (p>0.05) between ground-truth and automated prostate segmentation. Group mean values are presented in Table \ref{tab:table4} and Table \ref{tab:table5}.

%---------------------------------Table 4 ----------------------------------

\begin{table}
	\caption{Values obtained in PG, CZ and PZ regions of interest using ground truth and predicted masks. “mask” and “predicted” in the column headers refer to the ground truth and model output, respectively. T-test results between ground truth and predicted masks for PG, CZ, and PZ are reported in the right part of the table (based on segmentation results for the IMs using magnitude images as input).}
	\centering
	\resizebox{\columnwidth}{!}{%
	\begin{tabular}{l|cccccc|ccc}
		\toprule
		\multicolumn{6}{c}{\hspace{3cm}Pixel values (average ± standard deviation)} &
		\multicolumn{4}{c}{\hspace{2cm}T-test (significance at p<0.05)}\\
		\cmidrule(r){2-7}
		\cmidrule(r){8-10}
		Map & PG mask & PG predicted & CZ mask & CZ predicted & PZ mask & PZ predicted & PG P-value & CZ P-value & PZ P-value   \\
		\midrule
		SWS & 1.28±0.39 & 1.35±0.4 & 1.25±0.32 & 1.36±0.33 & 1.39±0.22 & 1.28±0.22 & 0.87 & 0.81 & 0.83 \\
		mag & 33.1±10.46 & 36.56±11.05 & 29.6±7.85 & 34.03±8.31 & 43.21±7.2 & 39.7±7.29 & 0.83 & 0.78 & 0.85\\
		$\varphi$ & 0.61±0.19 & 0.63±0.19 & 0.64±0.17 & 0.66±0.17 & 0.58±0.09 & 0.56±0.09 & 0.91 & 0.88 & 0.87\\

		\bottomrule
	\end{tabular}
	\label{tab:table4}
}
\end{table}

%---------------------------------Table 5 ----------------------------------

\begin{table}
	\caption{Values obtained in PG, CZ and PZ regions of interest using ground truth original and predicted masks. “mask” and “predicted” in the column headers refer to the ground truth and model output, respectively. T-test results between ground truth and predicted masks for PG, CZ and PZ are also reported in the right part of the table (based on segmentation results from UM and input images of mag).}
	\centering
	\resizebox{\columnwidth}{!}{%
	\begin{tabular}{l|cccccc|ccc}
		\toprule
		\multicolumn{6}{c}{\hspace{3cm}Pixel values (average ± standard deviation)} &
		\multicolumn{4}{c}{\hspace{2cm}T-test (significance at p<0.05)}\\
		\cmidrule(r){2-7}
		\cmidrule(r){8-10}
		Map & PG mask & PG predicted & CZ mask & CZ predicted & PZ mask & PZ predicted & PG P-value & CZ P-value & PZ P-value   \\
		\midrule
		SWS & 1.28±0.39 & 1.29±0.39 & 1.25±0.32 & 1.32±0.3 & 1.39±0.22 & 1.27±0.24 & 0.9 & 0.89 & 0.81\\
		mag & 33.1±10.46 & 34.1±10.6 & 29.6±7.85 & 30.54±7.0 & 43.21±7.2 & 39.6±8.05 & 0.95 & 0.95 & 0.85\\
		$\varphi$ & 0.61±0.19 & .62±0.19 & 0.64±0.17 & 0.66±0.16 & 0.58±0.09 & 0.57±0.11 & 0.99 & 0.92 & 0.95\\

		\bottomrule
	\end{tabular}
	\label{tab:table5}
}
\end{table}

\section{Discussion}
\label{sec:others}
Our study shows that Dense U-net segmentation of prostate zones based on MRE data allows automated tabulation of quantitative imaging markers for the total prostate and for the central and peripheral zones. 

\subsection{IMs versus UM}
\begin{figure}[H]
\centering

\includegraphics[width=15cm, height=18cm]{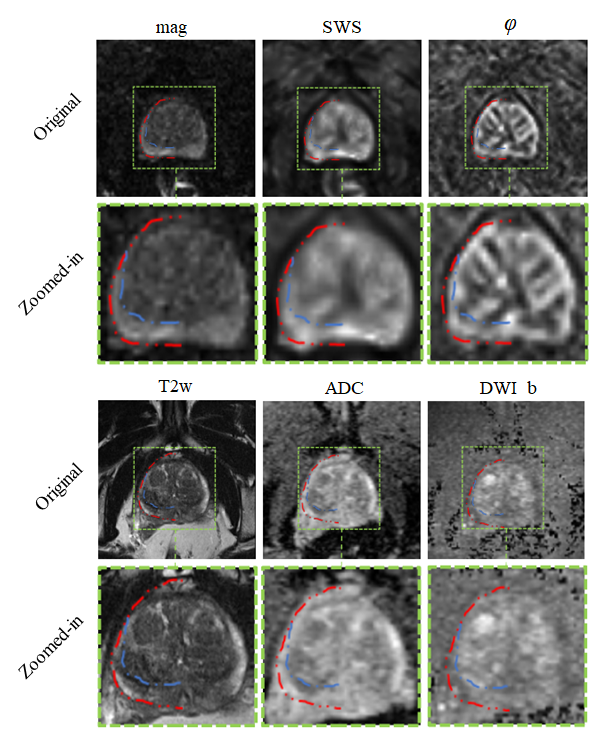}
\caption{Different maps of the same patient. Second and forth rows show zoomed-in images of the first and third rows, respectively. Dotted red and blue lines highlight part of the segmentation for prostate and PZ, respectively and the rest of the delineation was omitted to show the actual pixel intensities.}
\label{fig:9.png}
\end{figure}

Our results show that IMs performed excellent across all maps and sequences with high values for DS and low values for HD. In our experiments, segmentation was most reliable when we used T2w and MRE magnitude images, which provide sufficiently rich anatomical details for automated prostate segmentation while quantitative parameter maps such as SWS, $\varphi$ and ADC lack those details. Figure \ref{fig:9.png} depicts a variety of maps in a patient demonstrating that anatomy of prostate boundaries is well preserved on T2w and MRE magnitude images while it is less clearly visible on ADC, DWI\_b, SWS and $\varphi$ maps.

For IMs, we found no significant difference between DS of MRE and DS of MRI in PG and PZ while, in CZ, DS of MRE was higher than that of MRI. This may be explained in part by blurring due to larger slice thickness in MRI (3mm) than MRE (2mm). Furthermore, MRI slice volumes covered the entire prostate gland, the seminal vesicles and the periprostatic tissues while MRE volumes were solely focused on the prostate gland given the smaller slice thickness. 

In contrast to IMs, the UM could process any input combination of MRI/MRE maps without a need for retraining or fine-tuning the network. Similar to IMs, the UM also favored input combinations with rich anatomical detail, such as T2w and magnitude images. In all experiments, HD of the UM ranged from 1.15 to 5.29 mm, which is significantly inferior to IMs. Moreover, by showing the changes in DS and HD due to all possible input combinations for IMs and UM, Figure \ref{fig:7.png} illustrates that both IMs and the UM had decent performance while IMs were slightly better. Nevertheless, given the robustness of the UM combined with decently good segmentation results, it is recommended to primarily apply IMs and to use the UM as a second opinion for automated prostate segmentation.

\subsection{Quantitative results}
Exploiting MRE magnitude images for automatic segmentation instead of high-resolution in-plane T2-weighted MRI had the benefit of not requiring image registration and making full use of the anatomic and viscoelastic information contained in a single MRE dataset. This potentially stabilizes quantitative parameter extraction as any co-registration artifacts are avoided. As the ultimate proof of valid segmentations, viscoelasticity values averaged within volumes of PG, CZ, and PZ obtained from CNNs were not different from those obtained with manually segmented masks (see supplemental information). Thus, we here for the first time used the information of the magnitude signal in prostate MRE, showing that is was fully sufficient for accurate segmentation, which may greatly enhance MRE of the prostate in the future. 

\subsection{Challenges}
Figure \ref{fig:8.png} shows a case where the model failed to achieve accurate segmentation. Inaccuracies appear to be attributable to under-segmentation and discontinuity. Under-segmentation is visible in both the entire prostate gland (first row) and the CZ (last row), where the model did not properly locate and delineate boundaries. Discontinuity can be seen in the PZ (middle row), where the model resulted in a mask with several unconnected neighboring areas. Many factors can contribute to inaccurate segmentation, including boundary ambiguity, partial volume effects, and tissue heterogeneity. Therefore, radiologists typically use 3D information, which is subjectively interpolated by eye to the ambiguous image slice. However, even including adjacent slices for training in a 2.5-D approach \cite{RN162} or use of full 3D models does not necessarily lead to better segmentation performance due to partial volume effects \cite{RN118} .

\subsection{Future directions}
We will implement the proposed CNN-based segmentation on a server, which is currently used for MRE data processing (\url{https://bioqic-apps.charite.de}), in order to make it available to the research community. Our ultimate aim is to accomplish fully automated tabulation of prostate SWS and $\varphi$ values based on multifrequency MRE data. Once installed, the Dense U-net will be trained using other sets of MRE data acquired with other scanners and MRE sequences in order to generalize its applicability for prostate segmentation. In a next step, we aim at automated segmentation of other organs including the liver, kidneys, and pancreas based on MRE magnitude images. 

\subsection{Limitations}
Our study has limitations, including the small number of patients, which is attributable to fact that we performed a proof-of-concept exploration of CNNs for MRE-based prostate segmentation. For further improvement of segmentation quality and generalization to other domains, future studies should use multicenter MRI/MRE data acquired with different imaging protocols. Finally, we assembled all data for training in a way that avoids image alignment and registration procedures. While this approach ensured robust results based on single sets of data, we cannot rule out that combinations of MRI and MRE images (e.g., magnitude and T2w) would have resulted in slightly better DS and HD scores. 

\subsection{Summary}
Magnitude images of prostate MRE were used for automated segmentation of prostate subzones based on trained CNNs. As such, MRE data provide all information needed for extraction of viscoelasticity parameters and for delineation of the prostate regions for which those values are of interest for tabulation and automated classification of suspicious prostate lesions. Dense U-net achieved excellent segmentation results using both IMs and the UM and yielded MRE parameters that were not different from ground truth. Compared with standard image segmentation based on T2w images, MRE magnitude images proved to suffice, as demonstrated by excellent Dice scores and Hausdorff dimension results. Quantitative maps of multiparametric MRI including those of DWI and viscoelasticity did not provide adequate anatomic information for learning-based prostate segmentation. Prostate MRE combined with Dense U-nets allows tabulating quantitative imaging markers without manual analysis and independent of other MRI sequences and thus has the potential to contribute to imaging-based PCa detection and classification.

\section{Acknowledgments}
This work was funded by the German Research Foundation (GRK2260, BIOQIC; SFB1340, CRC Matrix in Vision).

\section{Conflict of interest declaration}
The authors declare no competing interests. MD is European Society of Radiology (ESR) Research Chair (2019–2022), and the opinions expressed in this article are the author’s own and do not represent the view of ESR. Per ESR guiding principles, the work as Research Chair is on a voluntary basis, and only travel expenses are remunerated.

\bibliographystyle{abbrv}
\bibliography{references}  %%% Uncomment this line and comment out the ``thebibliography'' section below to use the external .bib file (using bibtex) .





% \end{thebibliography}

\end{document}